
\magnification=1200

\font\tit=cmbx10 scaled\magstep2

\font\abst=cmti9

\font\abstb=cmbx9

\font\addrt=cmti9

\font\aun=cmbx10


\font\ridbf=cmr9

\font\rauth=cmcsc10

\font\tas=cmti9


\topskip=1truecm
\abovedisplayskip=15truept plus 5truept minus 5truept
\belowdisplayskip=15truept plus 5truept minus 5truept
\newskip\myskip
\myskip=40truept
\clubpenalty=1000
\widowpenalty=1000
\hsize=6truein
\vsize=8.5truein
\newcount\who
\who=0
\rm
\def\centra#1{\vbox{\rightskip=0pt plus1fill\leftskip=0pt #1}}
\def\centrb#1{\vbox{\rightskip=0pt plus1fill\leftskip=\myskip #1}}
\def\titolo{}
\def\frac#1#2{{#1\over #2}}

\def\title#1{\vfill\eject
\ifodd\pageno\else\def\firstauthor{}\null\vfill\eject\fi
\firstp=\pageno
\def\speak{}
\parindent=0pt
\baselineskip=8truept
\line{\hfill\vbox{\tas\spaceskip=0pt plus1fil
\hbox to 62mm{Proceedings of the XI Italian Conference on}
\hbox to 62mm{General Relativity and Gravitational Physics}
\hbox to 62mm{Trieste (Italy), September 26--30, 1994}
\hbox to 62mm{\copyright 1995 World Scientific Publishing Company}}}

\baselineskip=20truept
\bigskip\medskip
\centra{\tit #1}
\bigskip\baselineskip=12pt\def\titolo{#1}}

\def\oneauthor#1#2{\bigskip\centrb{\aun #1}\medskip
\centrb{\addrt #2}\global\def\firstauthor{#1}}


\long\def\support#1#2{\footnote{}
{\noindent\line{\hbox to 6truecm{\hrulefill}\hfill}
\hbox to 20truept{\hfill$^{#1}$\ }{\abst #2}}}

\parskip=5pt
\long\def\summary#1{\bigskip
\vbox{\par\leftskip=\myskip\rightskip=0truept\baselineskip=11pt
\noindent{\abstb Summary:} \abst #1 \bigskip\centrb{\speak}}
\parindent=20truept
\write1{\string\record{\titolo}{\folio}{\firstauthor}}
}

\def\section#1#2{\who=1\bigskip\medskip\goodbreak{\bf\noindent\hbox to
20truept{#1.\hfil}#2}\nobreak
\medskip\nobreak\who=0}
\def\acknow{\bigskip\medskip\goodbreak{\bf Acknowledgements}\nobreak
\medskip\nobreak}
\def\subsection#1#2{\ifnum\who=0\bigskip\goodbreak\else\smallskip\fi
{
\noindent\hbox to 25truept{\ridbf #1.\hfil} \ridbf #2}\nobreak
\ifnum\who=0\medskip\fi\nobreak}
\def\references{\bigskip\medskip\goodbreak{\bf\noindent\hskip25truept
References}\nobreak
\medskip\nobreak
\frenchspacing\pretolerance=2000\parindent=25truept}

\def\paper#1#2#3#4#5#6{\item{\hbox to 20truept{[#1]\hfill}}
{\rauth #2,} {\it #3} {{\bf #4},} #5 (#6)\smallskip}
\def\book#1#2#3#4#5{\item{\hbox to 20truept{[#1]\hfill}} {\rauth #2,}
{\it #3}, #4 (#5)\smallskip}
\def\booknn#1#2#3#4#5{\item{\hbox to 20truept{\hfill}} {\rauth #2,}
{\it #3}, #4 (#5)\smallskip}
\def\papernn#1#2#3#4#5#6{\item{\hbox to 20truept{\hfill}}
{\rauth #2,} {\it #3} {{\bf #4},} #5 (#6)\smallskip}
\def\paperibidem#1#2#3#4#5#6#7#8#9{\item{\hbox to 20truept{[#1]\hfill}}
{\rauth #2,} {\it #3} {{\bf #4},} #5 (#6); ibid. {{\bf #7},} #8 (#9)
\smallskip}
\def\bookpre#1#2#3{\item{\hbox to 20truept{[#1]\hfill}} {\rauth #2,}
(#3)\smallskip}

\newcount\firstp

\headline={\ifnum\pageno=\firstp\line{\hfill}\else
\ifodd\pageno\vbox to 12pt{\line{\hfill{\it\titolo}\hfill\folio}%
\vfill\hrule}\else%
\vbox to 12pt{\line{\folio\hfill{\it \firstauthor}\hfill}\vfill\hrule}%
\fi\fi}

\footline={\ifnum\pageno=\firstp\line{\hfill\folio\hfill}\else%
\line{\hfill}\fi}

\title{Relativity Experiments in the Solar System}

\oneauthor{Giacomo Giampieri}{Jet Propulsion Laboratory,
California Institute of Technology,
4800 Oak Grove Drive, Pasadena, CA 91109.}

\summary{Recent theoretical works on alternative metric theories of gravity
give greater significance to solar-system tests of General
Relativity. In particular, it is suggested that the post-Newtonian
parameter $\gamma$ ought to be determined with great precision in
order to discover possible preferred frame effects or
effects related to a tensor-scalar theory of gravitation. In this
work we focus on a future experiment, based on Doppler tracking of
an interplanetary spacecraft near superior conjunction.

We simulate, under certain assumptions, the frequency of the
signal received from a spacecraft equipped with the ultimate radio
capabilities. The geometrical optics formalism adopted
makes it particularly easy calculating the Doppler effect caused by
a spherically symmetric refractive medium. This case includes
both the post-Newtonian gravitational effect and a spherical solar
corona model. Stochastic fluctuations in the plasma electron
density, along with other sources of noise, were also considered in
simulating the data.

Applying a least squares fit to the simulated Doppler data, we are
able to determine $\gamma$ with a precision of $5\times10^{-4}$, a
factor four better than the present value. The availability of a
two-frequencies link to the spacecraft and back would allow an
improvement by at least an order of magnitude. Our results seem to
indicate that measuring the time derivative of the Shapiro
ranging curve is more  promising than the ranging technique
itself, although additional work comparing the various
methods is needed.}

\section{1}{Introduction}

The Parametrized Post-Newtonian (PPN) formalism is the standard
framework for performing experiments and formulating
gravitational theories in the weak field and slow motion limit,
appropriate to the solar system.
In fact, in this limit, the spacetime metric predicted by nearly
every  metric theory of gravity has the same structure. In other
words,  the metric can be expanded about the Minkowskian metric
through dimensionless gravitational potentials of varying degrees
of smallness. Metric theories differ one from another only in the
numerical values of the coefficients that appear in front of the
metric potentials. The PPN formalism prescribes the values of
these parameters for each particular theory under study.
Usually,
we consider only ten of these parameters, chosen in such a way
that their values indicate general properties of the metric theories
of gravity. Thus, the comparison of these theories with each other
and with the experiments is straightforward in this framework [1].

After almost eighty years since General Relativity (GR) was born,
we can conclude that Einstein's theory has survived every test.
This rare longevity, along with the absence of any adjustable
parameters in the theory (namely two of the ten PPN parameters
are equal to one, all others disappear), far from being
satisfactory,  seems to be the cause of a lasting search for a
possible fatal  discrepancy.
Remarkably, as an upshot of these efforts most alternative theories
have been put aside. Only those theories of gravity flexible enough
to accommodate the resulting constraints have survived, the
protective belt being provided by the free parameters and the
coupling constants of the theory.

The Brans-Dicke (BD) theory is certainly the most famous among
the alternative theories of gravity. It contains, besides the metric
tensor ${\bf g}$, a scalar field $\phi$ and an arbitrary coupling
constant $\omega$, related to the PN parameter $\gamma$ by
$$\gamma={1+\omega\over 2+\omega}\, .\eqno(1)$$
The present
limit on $\gamma$ (see eq.\ (3) below) gives the
very tight bound on the coupling constant $\omega>500$. In  addition,
many Scalar-Tensor (ST) theories, generalizing the  original BD theory,
have been proposed in which the coupling  function depends on the value
of the scalar field,  $\omega=\omega(\phi)$. These theories are
considered as the  most interesting alternatives to GR, due to recent
developments in  cosmology (e.g.\ inflationary models) and
elementary-particle  physics (e.g.\ string theory and Kaluza-Klein
theories).  As recently pointed out by various authors (see [2,3]),
a large class of ST theories contain an
attractor  mechanism toward GR in a cosmological sense. If this is what
actually happened, then today we can expect, from eq.\ (1), a small
deviation of $\gamma$ from the GR value $\gamma=1$. The
present level of this deviation is estimated to be in the interval
$$\delta\gamma \approx 10^{-7} - 10^{-5}\eqno(2)$$
the exact value depending on the particular ST theory under
investigation [3].

\section{2}{A test based on Doppler technique}

The previous arguments provide strong motivations for experiments
able to push beyond the present empirical accuracy on $\gamma$,
which is [4]
$$\gamma= 1.000 \pm 0.002\, ,\eqno(3)$$
where the quoted uncertainty is twice the standard deviation.

A number of advanced experiments and space missions have been
proposed, in order to improve this level of accuracy by at least
two orders of magnitude (see, for example, [5]).
In addition to these, the interplanetary mission CASSINI, which will be
launched in October 1997,
offers a new possibility for a test based on a free-flying
spacecraft.  The instrumentation will include a coherent Doppler
capability  with simultaneous tracking at X ($\sim8.4\hbox{\sl GHz}$)
and Ka ($\sim33\hbox{\sl GHz}$)  radio bands. Since CASSINI is the first
deep space mission to  implement Ka-band communications and
tracking,
a significant  improvement over the best published results using
S-band  ($\sim2.3\hbox{\sl GHz}$) and X-band is expected.

In Doppler experiments, a radio link is transmitted from the Earth
to the spacecraft, coherently transponded and sent back to the
Earth, where its frequency is measured with great accuracy,
thanks to the availability of hydrogen masers at the stations used
for interplanetary telecommunications. Comparing the transmitted and
received frequencies, one can measure the Doppler shift
$$y={\Delta \nu\over \nu_0}={1\over c}{dl\over dt}\, ,\eqno(4)$$
where $l$ is the overall optical distance (including diffraction
effects) traversed by the photon in both directions.
This technique has various applications, from the estimate of the
gravity field of planetary systems (see [6,7] for recent examples),
to the study of interplanetary plasma [8,9], and the attempts to
detect low frequency  gravitational waves [10-12]. Interplanetary Doppler
tracking near solar conjunction, as a tool for testing relativistic
effects, has been discussed previously in [13].

Besides the ordinary Doppler effect, of order $v/c$, two
relativistic  corrections of order $(v/c)^2$ contribute to the
Doppler signal: the  gravitational shift, which occurs when the
gravitational potential at  the source and the receiver is different,
and the second order effect,  well known in special relativity. Both
effects have been tested using  a space-borne hydrogen maser carried
aboard a sub-orbital rocket, allowing a test of  Einstein's prediction to
about $10^{-4}$ [14].
A different, third order relativistic effect is observable
in interplanetary Doppler measurements and, indeed, is implicitly
incorporated in the orbit determination numerical codes [15].
In experiments near superior conjunction, with an impact parameter
$b$, the ray suffers a deflection
$$\delta = {4 M_\odot\over b} = 8\cdot10^{-6}{R_{\odot}\over b}\, ,\eqno(5)$$
where $M_\odot=1.5~\hbox{\sl km}$ is the
gravitational radius of the sun.  This deflection changes the velocity
component along the line of  sight and will produce an additional
Doppler effect of order
$$v\,\delta = {4 M_\odot\over b} v \approx
8\cdot10^{-10}{R_{\odot}\over b}\, .\eqno(6)$$
Here and in the following we have taken $G=c=1$.
Because of the different technique, this can be considered as an
additional test of GR, related to the time-delay effect in radar
propagation first pointed out by Shapiro [16].
Since in a
generic metric theory the expressions (5) and
(6)  are multiplied by $(1+\gamma)/2$, the experiment amounts to
a measurement of $\gamma$.
The current stability of hydrogen masers ($10^{-15}$ or better)
would allow a significant improvement in the accuracy of
$\gamma$, assuming that the effect of the solar corona can be
overcome using dual-frequency capability.
Smaller effects, such as the gravito-magnetic force acting on
photon propagation, can also be studied with a `differential'
Doppler technique [13].

Before the introduction of multi-link tracking signal, including
links at Ka band, the technique of measuring the relativistic
Doppler shift caused by solar gravity was not competitive with the
ranging technique. Our initial assessment is that the Doppler
technique will be more powerful than ranging for advanced
experiments like CASSINI. The purpose of our work is to
quantify that expectation.
The plan of the paper is as follows.  In sec.3 we derive the
gravitational Doppler effect, using standard geometrical optics
formalism. The same formalism gives, in sec.4.1, the steady-state
contribution from the solar plasma. Sec.4.2 describes how to simulate
a real experiment. The simulated data are then analyzed in sec.5, in
order to predict the accuracy in the determination of the PN parameter
$\gamma$, when single and dual-frequencies links are available. Some
of the limits of our analysis are then presented and discussed.
Finally, in sec.6, we consider the Doppler technique in connection
with two well-known alternatives, namely
ranging and Differential VLBI ($\Delta$VLBI).

\section{3}{Gravitational Doppler shift}

It is well known that null geodesics of a static spacetime are curves of
extremal coordinate time interval between the two end-points [17].
In particular, null geodesics for the post-Newtonian metric
$$ds^2 = -(1-2U) dt^2 +(1+2\gamma U)\delta_{ij} dx^i dx^j$$
generated by a massive central body ($U=U(r)=M_\odot/r$) can be
derived from the variational principle
$$\delta \int_\Gamma \mu\ d\lambda = 0\, ,$$
where
$$\mu = 1 + (1+\gamma){M_\odot\over r}$$
is the refractive index, $d\lambda =\sqrt{\delta_{ij} dx^i dx^j}$ is
the geometrical arc-length (in flat space), and the spatial path
$\Gamma$ is to be varied with fixed end-points.
The photon's trajectory is thus obtained by integrating the
ray-equation
$${d~\over d\lambda}\left(\mu \vec{p}\,\right) = \vec\nabla\mu \,,\eqno(7)$$
where $\vec{p}$ is the unit vector tangent to the path.
Since we are dealing with the first PN effect, we can integrate
eq.\ (7)  along the zero-th order trajectory, i.e.\ the straight
line from the  source to the receiver.
\bigskip
\bigskip\noindent
{\abst Fig.\ 1: The deflection geometry.}
\bigskip
We choose the geometry as in fig.1; $z=0$ is the plane through the
sun, the earth and the spacecraft at any given time, and the
unperturbed ray is directed along the {\sl x}-axis, so that
$\vec{r}=(\lambda,b,0)$. We get, in the limit $\ell_0, \ell_1
\gg b$,
$$\Delta \vec{p}  =  \int\limits_{-\infty}^{+\infty} \vec\nabla
\mu ~d\lambda = -{2(1+\gamma)M_\odot\over b}\, \hat{y}\, .\eqno(8)$$
Projecting eq.\ (8) along the {\sl y}-axis we recognize the
usual  expression for the angular deflection of light, so that we will
adopt  the following helpful notation
$$\Delta \vec{p} \equiv \Delta\theta_{gr}(b)\, \hat{y} \,,
\qquad\qquad
\Delta\theta_{gr}(b) = -{2(1+\gamma)M_\odot\over b}\, .\eqno(9)$$
The one-way Doppler shift is given by the  invariant expression
$${\nu_1\over \nu_0}={{\bf v}_1\cdot{\bf p}({\bf x}_1)\over
{\bf v}_0\cdot{\bf p}({\bf x}_0)}\eqno(10)$$
where ${\bf x}_i$ and ${\bf v}_i$ are,
respectively, the position and velocity of the receiver (emitter).
To  lowest order the photon momentum is ${\bf p} =
(1,1,0,0)$, which  gives rise, when substituted in eq.\ (10),
to the ordinary special  relativistic effect. This effect changes over
the Keplerian time  scale $\sqrt{\ell^3/M_\odot}\approx \ell/v$. The
gravitational  contribution to the Doppler signal comes instead from
the first  order correction $\delta{\bf p}$ to
${\bf p}$, and it changes over  the time scale $b/v$.
These corrections are needed at the  end-points. Since we are dealing
with a static metric, the time  component $\delta p^0=0$. The space
components are obtained  from eq.\ (8) after imposing the
aiming condition, i.e.\ the passage  through ${\bf x}_0$ and ${\bf x}_1$:
$$\int\limits_{-\ell_0}^{\ell_1} \delta\vec{p}\ d\lambda=0\, .\eqno(11)$$
The result is
$$\eqalignno{\delta\vec{p}\,({\bf x}_0) &=
-\Delta\theta_{gr}(b)
{\ell_1\over\ell_0+\ell_1}\, \hat{y}&(12)\cr
\delta\vec{p}\,({\bf x}_1) &=
+\Delta\theta_{gr}(b)
{\ell_0\over\ell_0+\ell_1}\, \hat{y}\, .&(13)\cr}$$
Finally, substituting eqs .(12), (13) in
eq.\ (10) we get the gravitational  Doppler shift
$$y_{gr}(t)= -\left(v^y_{1} \ell_0 + v^y_{0}
\ell_1\over\ell_0+\ell_1\right) \Delta\theta_{gr}(b)\, .\eqno(14)$$
The first factor in eq.\ (14) depends on the spacecraft
and Earth's orbits
(according to fig.1, with $v^y$ we indicate the velocity orthogonal
to the line of sight), while the second factor depends on time
through the impact parameter $b$, which, near conjunction,
changes almost linearly in time. Of course, the round trip, or
two-way, Doppler signal is simply obtained multiplying eq.\ (14)
by two.

Figure 2 shows an example of this signal, as a function of time,
when the spacecraft is on a circular orbit around the sun, with
Mercury's  orbital radius. We have assumed $\gamma=1$ in this case,
and chosen  the origin of time at closest approach ($t_c=0$).
Moreover, the orbital plane is inclined by
$\sim1^0$ with respect to the ecliptic, so that the ray grazes the
sun limb when $b$ is minimum (in other words, $b(t_c)\simeq R_\odot$).
Note also that at closest
approach the special relativistic effect, not shown here, has its
minimum, since at this epoch the velocities are orthogonal to the
photon's path, and we have only a second order effect.
\bigskip
\bigskip\noindent
{\abst Fig.\ 2: The general relativistic Doppler shift due to the solar
monopole. The inclination $I$ between the ecliptic plane and the
spacecraft orbital plane is large enough to prevent the solar
occultation, i.e. $I\simeq 1^0\Rightarrow b(t_c)\simeq R_\odot$.
The experiment is assumed to occur near the point where the
satellite has the largest distance from the ecliptic.}
\bigskip

\section{4}{Plasma effects}

The solar corona plasma is the main source of noise in a radio
experiment near conjunction.
In general, we can express the Doppler signal as
$$y=y_{gr} + y_{pl} + y_n\, ,\eqno(15)$$
where $y_{pl}$ is the dominant plasma noise, and $y_n$ contains all
non-dispersive sources of noise (clock, receiver, etc.), along with
a minor dispersive contribution from the atmosphere.

The plasma contribution $y_{pl}$ to the fractional
frequency change (4) is related to the change in the optical
path
$$\Delta l = {N_e e^2\over 2\pi m_e \nu_0^2}\, ,$$
where $e$ is the electron's charge, $m_e$ its mass, and $N_e$ the
total columnar content along the beam, $N_e =\int\! n_e \, d\lambda $.
Therefore, in order to calibrate for the plasma term, we should know
the electron density along the path. We start by decomposing
the electron density $n_e$ in a static, spherically
symmetric part $\langle n_e \rangle (r)$
plus a  fluctuation $\delta n_e$, i.e.
$$n_e(\vec{r},t) = \langle n_e \rangle (r) + \delta n_e(\vec{r},t)
\, .\eqno(16)$$
As a matter of fact, the steady-state behavior is
reasonably well known, and we can use one of the several plasma models
found in the literature [18-21]. To be
more explicit, we will refer to two particular models,
namely\footnote{$^1$}{\abst We do not consider here a correction factor
which depends on the heliographic latitude.}
$$\langle n_e\rangle(r)=\cases{
\displaystyle{\left[\left({2.99\over\eta^{16}}
+{1.55\over\eta^6}\right) \times10^8
+{3.44\times 10^5\over\eta^2}\right]~\hbox{\sl cm}^{-3}}
&\qquad\hbox{model (a)}\cr\cr
\displaystyle{\left[{2.39\times10^8\over\eta^6}+
{1.67\times 10^6\over\eta^{2.3}}\right]~\hbox{\sl cm}^{-3}}
&\qquad\hbox{model (b)}\cr\cr}\eqno(17)$$
where $\eta\equiv r/R_\odot$. The reason why we consider both models
is  two-fold: first, repeating the same analysis with two different
electron distributions, we can test the model-dependency of the
results  presented here. The second reason is more subtle; in the
following  section we will describe a simulation and a fit of the
real data, so  that, if we want to obtain meaningful estimates for
$\delta\gamma$,  we cannot use the same model in both operations.
More on this later.

\subsection{4.1}{Steady state coronal plasma}

We will now determine the Doppler effect due to the solar corona
plasma for a model of the form (17a) or (17b).
More precisely, we  consider a generic plasma model of the form
$$\langle n_e \rangle(r) = \sum\limits_k{\alpha_k\over\eta^{\beta_k}}\, .$$
According to the electromagnetic law of phase propagation through a plasma
me\-dium, the  refraction index $\mu$ can be expressed in terms of the plasma
frequency $\nu_p$ and the carrier frequency $\nu_0$ as
$$\mu = \sqrt{1-\left(\nu_p\over\nu_0\right)^2} \simeq
1-{\langle n_e \rangle e^2\over 2\pi m_e \nu_0^2} \,.\eqno(18)$$
We can now insert eq.\ (18) in the eikonal equation (7), and
integrate along the unperturbed trajectory $\vec{r}=(\lambda,b,0)$,
exactly as we did in the previous section for the relativistic
$\mu$ (see fig.1). In the limit $\ell_0,\ell_1 \gg b$ we
easily find
$$\Delta\vec{p} = \Delta\theta_{pl}(b)\, \hat{y}\, ,$$
where now
$$\Delta\theta_{pl}(b) = {e^2\over 2\pi m_e \nu_0^2} \sum\limits_k
\alpha_k \beta_k \left(R_\odot\over b\right)^{\beta_k}
B\left({1+\beta_k\over 2},{1\over 2}\right)\,,\eqno(19)$$
and $B(x,y)$ is the Euler's Beta function.
Comparing $\Delta\theta_{pl}$ with $\Delta\theta_{gr}$ we notice the
opposite sign - gravity bends the ray outwards, plasma inwards - and
the different dependence on $b$, plasma effect being steeper.
The plasma deflection as a function of the
solar offset $b$ is shown in fig.3. In this case, we adopted for the
steady-state model the first of eqs.\ (17), and considered S, X,
and  K-bands radio frequencies. The absolute value of
the GR bending is also shown. The same analysis for the second plasma
model (eq.\ (17b)) does not yield any appreciable difference, so
we shall  omit it here.
\bigskip
\bigskip\noindent
{\abst Fig.\ 3: The deflection angle as a function of the impact parameter
$b$. The solid line is the plasma contribution, while the dashed line
gives the absolute value of the general relativistic deflection,
given by eq.\ (9) with $\gamma=1$.}
\bigskip
Imposing the boundary condition (11), as we did for the
gravitational  effect, we can now obtain $\delta\vec{p}\,({\bf x}_i)$.
Inserting  these in eq.\ (10) we
eventually get the steady-state part of the
solar plasma Doppler shift.
The details of this
calculation have already been carried out in the previous section.
The result is consequently
analogous to the gravitational Doppler shift, i.e.
$$y_{pl}(t)= - \left(v^y_{1} \ell_0 + v^y_{0} \ell_1\over\ell_0+
\ell_1\right) \Delta\theta_{pl}(b)\qquad\qquad
\hbox{(steady-state plasma)}\eqno(20)$$
Note from eqs.\ (14) and (20) that the first factor is
the same as in $y_{gr}$. Both effects are time dependent through the
quantities $\ell_0,\ell_1,\vec{v}_0,\vec{v}_1,$ and $b$.

\subsection{4.2}{Towards a real experiment}

In the previous paragraph we have considered a spherical and
static corona
model, and derived the resulting Doppler shift, according to the law
of geometrical optics. Unfortunately, the true electron density
(16)  contains also the fluctuations $\delta n_e$, which require
particular  attention. In fact,
these fluctuations are carried along with the solar wind speed
$V\simeq 400~\hbox{\sl km/sec}$, so that their
spatial scale is  $V\tau\simeq R_\odot/2$, where
$\tau\simeq1000~\hbox{\sl sec}$ is  the  temporal
scale. On the scale $b$ typical of the  gravitational Doppler effect,
one expects $\delta n_e$ to be of the same  order of magnitude as its
average [22], i.e.
$$\delta n_e(b,t) \approx \langle n_e \rangle(b)\,.$$
With the differential Doppler technique, introduced in [13],
the plasma contribution can be greatly reduced by
at least three orders of magnitude. However, this technique requires
a hydrogen maser aboard the spacecraft, which is beyond the scope of
the present study.

In conclusion, in order to be realistic we should
consider a stochastic signal of the same order of magnitude as the
the steady-state plasma effect $y_{pl}$, as given by
eq.\ (20).  Therefore, we have generated a random gaussian
noise of amplitude $y_{pl}(t)$, at each instant of time. For
simplicity, we did not take  into account the fact that this noise is
not white; although this  could be done quite easily, assuming for
example a Kolmogorov's spectrum, we  claim that this assumption does
not affect very much our results.

Finally, one must consider the non-plasma noise $y_n$ which
contributes to the overall signal (15). To be conservative, we
have assumed here
$$y_n\simeq 5\times 10^{-14}\eqno(21)$$
which takes into account the fact that the observations occur during
day time, although the noise level (21) could be pessimistic in
view, for example, of CASSINI specifications [23].

Summing up all these contributions, we can give a reasonable
description of the real data. These have been simulated in S, X, and
K-bands,  assuming the same geometrical configuration as in fig.2,
and adopting the solar corona model given by eq.\ (17a).
The simulated data are shown in  fig.4. A
superficial look to the K-band plot  could give an idea of the
expected improvement  over previous experiments, based on S and
X-band tracking.  Even a single K-band link could therefore
provide a significant estimate of the PN parameter $\gamma$. The
following section provides a tentative result for this expectation.
\bigskip
\bigskip\noindent
{\abst Fig.\ 4: The simulated data for the three bands. The axes and the
flyby's geometry are the same as in fig.2, where only the
gravitational Doppler shift was shown.}
\bigskip
\section{5}{Regression analysis}

As we have seen, the main difference between the gravitational and
the plasma Doppler effects is that the latter is much more steeper,
as a function of $b$ or $t$, than the former. This fact implies that
we can try to isolate the gravitational contribution with a
regression analysis. To be more explicit, given the $N$ data points
$y_i$ and  their variance $\sigma_i^2$, we want to minimize the
$\chi^2$ function
$$\chi^2=\sum\limits_{i=1}^N
\left(y_i-y_{\det}(t_i;\gamma,\alpha_1,\alpha_2,\dots)\over
\sigma_i\right)^2\,,$$
where $y_{\det}$ is the deterministic signal,
sum of the gravitational term (14), with $\gamma$
as a parameter,  plus the steady-state term (20), which
contains the  parameters $\alpha_1, \alpha_2,\dots$.
Thus
$$y_{\det}(t;\gamma,\alpha_1,\alpha_2,\dots) =
- \left(v^y_{1} \ell_0 + v^y_{0} \ell_1\over\ell_0+
\ell_1\right) \left(\Delta\theta_{gr} +
\Delta\theta_{pl}\right)\,.\eqno(22)$$
We have implicitly assumed that
the exponents $\beta_k$ are not varied by the fitting process. As we
anticipated, we have created the data $y_i$ using the
three-component model (17a), but it would not be fair to use this
information here, in the regression analysis.  In other words, in
order to take into account our ignorance about the `true' spherical
model that Nature has chosen, we will attempt a fit with the `wrong'
one, in this case the two-component model (17b), characterized
by $\beta_1=6, \beta_2=2.3$. Despite its formal importance, this
distinction is however irrelevant, as long as the models used are all
reasonable approximations one of the other.

We stress the fact that the orbital factor in the
relativistic Doppler signal (14), is {\it multiplied} by a
quantity of order $O(M_\odot/b)$, so that its Newtonian value suffices.
The analogous quantity in the time delay expression (see eq.\ (26)
below) is {\it added} to the relativistic correction, so that they are
both required at the same order. This constitutes a major
advantage of the Doppler method compared to the ranging technique.
Nonetheless, it is obvious that the orbital quantities appearing in
$y_{det}$ are affected by statistical errors, due for example
to  non-gravitational forces. We will discuss these orbital stochastic
effects, along with small variations resulting from dependence on the
PN parameter $\beta$, in a subsequent paper.

Before we proceed, we should address the problem of the optimal time
sequence to be used. Since the plasma, both the steady-state and the
stochastic terms, become more and more important approaching the
closest approach ($t=0$ in fig.4), we can decide to cut-off that
portion of the data, which introduces more noise than signal. Figure
5 represents the signal-to-noise ratio as a function  of
the cut-off impact parameter, indicated with $b_{cut}$.
Of course, dealing with the $\chi^2$ function avoids this
complication. In fact, the post-fit residuals give
$\delta\gamma/\gamma$ as a function of the minimum impact
parameter
$b_{cut}$ considered (fig.6). Comparing figures 5 and 6 we can
conclude that the accuracy with which we can determine $\gamma$
increases as we track closer to the sun, up to a certain point (in
the K-band case around  $b_{cut}\simeq 5 R_\odot$), after which plasma
effects begin to dominate and deteriorate the accuracy, so that we
should not push the  observations beyond this threshold.
\bigskip
\bigskip\noindent
{\abst Fig.\ 5: The Signal-To-Noise Ratio as a function of the
cut-off impact parameter $b_{cut}$. For each value of $b_{cut}$
only the first portion of the data, from the beginning of the
observations to the point where $b=b_{cut}$, are considered.}
\bigskip
\bigskip\noindent
{\abst Fig.\ 6: The fractional error
$\delta\gamma/\gamma$ as a function of the cut-off impact
parameter $b_{cut}$, where $b_{cut}$ has the same role as in
Fig.\ 5.}
\bigskip
In conclusion, from our regression analysis  we can infer that the
availability of K-band tracking could allow us to determine the PN
parameter $\gamma$ with a two-sigmas accuracy
$${\delta\gamma\over\gamma}\simeq 5\times 10^{-4}\,,\eqno(23)$$
a factor 4 better than the present value $2\times10^{-3}$, but still
far from  the optimal target quoted in sec.1 (see eq.\ (2)).
However, this sensitivity would be enough to detect possible effects
related to the existence of a preferred frame in the Universe. The solar
system would have, with respect to this reference frame, a velocity
$\sim10^{-3}$, which therefore represents a natural threshold for
possible effects at variance with GR.

It seems very difficult to go much beyond the result (23)
with a single link.  Even in K-band, there are limitations imposed by
plasma effects. However, since the plasma noise is dispersive, a
drastic improvement can be achieved when simultaneous multifrequency
links are available. For instance, if we track the spacecraft in X
and K bands, then from the two Doppler observables $y^K(t)$ and
$y^X(t)$ we can construct the quantity
$$\displaystyle{{y^K-\left(\displaystyle{{\nu_X\over\nu_K}}\right)^2
y^X\over 1-\left(\displaystyle{{\nu_X\over\nu_K}}\right)^2}} \simeq y_{gr} +
\displaystyle{{y_n^K-\left(\displaystyle{{\nu_X\over\nu_K}}\right)^2
y_n^X\over 1-\left(\displaystyle{{\nu_X\over\nu_K}}\right)^2}}\eqno(24)$$
which, in principle, does
not contain the frequency-dependent plasma noise. A  preliminary
regression analysis shows that this observable offers the opportunity
for a measurement of $\gamma$ with a precision at least an order of
magnitude better than eq.\ (23).

Given our simple simulation of the real data, these
estimates represent our best guess about the outcome of a
gravitational test based on  Doppler tracking of a spacecraft
equipped with the ultimate radio capabilities. However, before we
conclude, we would like to discuss the weak points in our
analysis. To start with,  we have
simulated our data without imposing any  strict obedience to an
actual proposed experiment; although the opportunity offered by the
CASSINI mission stimulated most of our work, we did not pay much
attention to the CASSINI expected orbit, error budget, etc. [23].
When necessary, however, we
always tried to assume realistic, if not pessimistic, numbers. A
more crucial caveat is that we neglected  some possible
complications, like the presence of the solar magnetic field, the
evolution of plasma properties over the duration of the experiment,
non gravitational accelerations, and so on.  In addition, our claim
that we can get rid of  plasma effects using multiple links is only
partially true. For instance, one must consider that two radio
signals at different frequencies traverse different paths, since
the bending (19) is frequency-dependent. As a
consequence, the plasma noise can not be completely removed from
eq.\ (24). However, we can predict how relevant is this
limitation. At closest approach, when the separation between the
two frequencies is bigger, we have
$$\Delta b_0 (b) \equiv b_0(X,b) - b_0(K,b) \approx
{0.1 R_\odot\over (b/R_\odot)^6}\,,\eqno(25)$$
where $b_0$ is the minimum distance of the ray from the sun (not to
be confused with the impact parameter $b$).
Comparing this separation with the typical length
scale of the plasma clouds, $V\tau\approx R_\odot/2$, we can conclude
that this is not much of a problem.  The result of this analysis is
summarized in fig.7, where the actual difference in the
closest approach distance of the K-band and X-band radio signals is
shown as a function of the impact parameter.
\bigskip
\bigskip\noindent
{\abst Fig.\ 7: The difference $\Delta b_0$ between the closest approach
distances in X and K bands, as a function of the impact parameter
$b$. The dotted line is the estimate (25).}
\bigskip
\section{6}{Future developments}

In the present work, we have considered a relativistic test based on
Doppler technique. Of course, Doppler is not the only technique
available when a free spacecraft is orbiting beyond the sun. Two
other well established experiments may also be performed at
conjunction, namely ranging experiments,
aimed to observe the Shapiro time delay, and
$\Delta$VLBI measurements of the deflection of light by
the solar mass.

The Shapiro time delay [16]
consists in the retardation of a light signal passing near a massive
body (the sun in our case). The round trip light time
(RTLT) from the Earth to the spacecraft
and back is easily obtained from eq.\ (7) as
$$\Delta t = \int \mu\, d\lambda \simeq \left(\Delta t\right)_{Newt} +
2(1+\gamma)M_\odot \ln\left(4 \ell_0 \ell_1\over b^2\right)\,.\eqno(26)$$
Unfortunately, we do not have direct access to the Newtonian signal $
\left(\Delta t\right)_{Newt}$ (this is somewhat related to the fact that
we are using a coordinate-dependent quantity),
so that what people do is a differential
measurement of the variations in RTLT as the spacecraft passes through
superior conjunction. This requires a good knowledge of the orbital
motion of the target relative to the Earth. For this reason, the best
result, quoted in eq.\ (3), has been obtained with the
Viking landers, anchored to the  surface of Mars [4],
while the more recent free-flying Voyager 2 test gave
$\delta\gamma\simeq 0.03$ [24].

Improvements in the accuracy of Very-Long-Baseline-Interferometry
(VLBI)
made it possible to measure differential angular positions to the level
of hundreds of microarcsec, making highly accurate measurements of
the deflection of light possible [25]. This experiment
is conceptually simple,  and does not involve secondary parameters
which
limit the confidence of  ranging experiments. A series of observations
yielded a value  ${1\over2}(1+\gamma)=1.000\pm0.001$ [26],
comparable to the Viking test of the Shapiro time delay.

In conclusion, every technique is based on
different concepts, and presents its own problems and peculiarities.
Therefore, deciding which one should be pursued during a given
mission is a difficult task. The project's management has to make a
decision based on the particular instrumentation available, on the
designated orbit, etc. It might also be advantageous to combine the
three techniques in an optimal way, keeping in mind the technical
difficulties involved in a multiple-task experiment.  Here, without
attempting such approach, we have just considered one particular
technique, and tried to define its possibilities. The comparison with
the others will be the object of future studies.

\acknow

The author is very grateful to John Anderson, John Armstrong, and
Bruno
Bertotti for their support and advice. He also thanks the Organizing
Committee of the XI Italian Relativity Meeting, and especially Mauro
Carfora, for their kind hospitality.
The research described in this paper was performed while the author
held
an NRC-NASA Resident Research Associateship at the Jet Propulsion
Laboratory, California Institute of Technology, under a contract with
the National Aeronautics and Space Administration.

\references

\item{\hbox to 20truept{[1]\hfil}}{\rauth C.M. Will}, {\it Theory and
Experiment in Gravitational Physics}, (Cambridge University Press, 1993)
\smallskip

\paper{2}{T. Damour and K. Nordtvedt}{Phys. Rev. Lett.}{70}{2217}{1993}

\paper{3}{T. Damour and K. Nordtvedt}{Phys. Rev.}{D48}{3436}{1993}

\paper{4}{R.D. Reasenberg, I.I. Shapiro, P.E. MacNeil,
R.B. Goldstein, J.C. Breidenthal, J.P. Brenkle, D.L. Cain,
T.M. Kaufman, T.A. Komarek and A.I. Zygielbaum}{Astrophys. Jour.}
{234}{L219}{1979}

\paper{5}{P.L. Bender, N. Ashby, M.A. Vincent and J.M. Wahr}
{Adv. Space Res.}{9}{113}{1989}

\paper{6}{J.K. Campbell and J.D. Anderson}{Astron. Jour.}{97}{1485}{1989}

\paper{7}{G. Schubert, D. Limonadi, J.D. Anderson, J.K. Campbell
and G. Giampieri}{Icarus}{111}{433}{1994}

\paper{8}{R. Woo}{Astrophys. Jour.}{201}{238}{1975}

\paper{9}{R. Woo, J.W. Armstrong, N.R. Sheeley, R.A. Howard,
M.J. Koomen, D.J. Michels and R. Schwenn}{J. Geophys. Res.}{90}{154}{1985}

\paper{10}{F.B. Estabrook and H.D. Wahlquist}{Gen. Rel. Grav.}{6}{439}{1975}

\paper{11}{B. Bertotti, R. Ambrosini, S.W. Asmar, J.P. Brenkle,
G. Co\-mo\-ret\-to, G. Giampieri, L. Iess, A. Messeri and H.D. Wah\-lquist}
{Astr. Astrophys. Suppl. Ser.}{92}{431}{1992}

\paper{12}{J.D. Anderson, J.W. Armstrong and E.L. Lau}{Astrophys. Jour.}
{408}{287}{1993}

\paper{13}{B. Bertotti and G. Giampieri}{Class. Quantum Grav.}{9}{777}{1992}

\paper{14}{R.F.C. Vessot and M.W. Levine}{Gen. Rel. Grav.}{10}{181}{1979}

\book{15}{T.D. Moyer}{JPL-NASA Technical Report}{32-1527}{1971}

\paper{16}{I.I. Shapiro}{Phys. Rev. Lett.}{13}{789}{1964}

\paper{17}{T. Levi-Civita}{Nuovo Cimento}{16}{105}{1918}

\paper{18}{G.L. Tyler, J.P. Brenkle, T.A. Komarek and A.I. Zygielbaum}
{J. Geophys. Res.}{82}{4335}{1977}

\paper{19}{D.O. Muhleman, P.B. Esposito and J.D. Anderson}{Astrophys. Jour.}
{211}{943}{1977}

\paper{20}{D.O. Muhleman and J.D. Anderson}{Astrophys. Jour.}{247}
{1093}{1981}

\paper{21}{M.K. Bird}{Space Sci. Rev.}{33}{99}{1982}

\paper{22}{J.W. Armstrong, R. Woo and F.B. Estabrook}{Astrophys. Jour.}
{230}{570}{1979}

\item{\hbox to 20truept{[23]\hfil}}{\rauth Cassini Project Mission Plan},
published as {\it JPL Document}, {\bf D-5564, Rev B} (1993)\smallskip

\paper{24}{T.P. Krisher, J.D. Anderson and A.H. Taylor}{Astrophys. Jour.}
{373}{665}{1991}

\paper{25}{E.B. Fomalont and R.A. Sramek}{Comm. Astrophys.}{7}{19}{1977}

\paper{26}{D.S. Robertson, W.E. Carter and W.H. Dillinger}{Nature}
{349}{768}{1991}

\bye